# Extracting Feelings of People Regarding COVID-19 by Social Network Mining

Hamed Vahdat-Nejad*, Fatemeh Salmani, Mahdi Hajiabadi, Faezeh Azizi, Sajedeh Abbasi, Mohadese Jamalian, Reyhane Mosafer

PerLab, Faculty of Electrical and Computer Engineering, University of Birjand, Iran

{vahdatnejad, salmani_fatemeh98, mahdihajiabadi, faezeh.azizi1995, sajedeh_abbasi, mohadesejamalian, reyhane.mosafer}@birjand.ac.ir

Hamideh Hajiabadi

Department of Computer Engineering, Birjand University of Technology, Iran

hajiabadi@birjandut.ac.ir

**Abstract:**

In 2020, COVID-19 became the chief concern of the world and is still reflected widely in all social networks. Each day, users post millions of tweets and comments on this subject, which contain significant implicit information about the public opinion. In this regard, a dataset of COVID-related tweets in English language is collected, which consists of more than two million tweets from March 23 to June 23 of 2020 to extract the feelings of the people in various countries in the early stages of this outbreak. To this end, first, we use a lexicon-based approach in conjunction with the GeoNames geographic database to label the tweets with their locations. Next, a method based on the recently introduced and widely cited RoBERTa model is proposed to analyze their sentimental content. After that, the trend graphs of the frequency of tweets as well as sentiments are produced for the world and the nations that were more engaged with COVID-19. Graph analysis shows that the frequency graphs of the tweets for the majority of nations are significantly correlated with the official statistics of the daily afflicted in them. Moreover, several implicit knowledge is extracted and discussed.

## 1. Introduction:

The COVID-19 pandemic has drastically changed lifestyles around the world. It has negatively affected employment, economy, and welfare in communities. In most nations, governments have imposed home quarantine, increased attention to personal hygiene, and the use of masks to prevent its further spread. Home quarantine has reduced physical communications between individuals which has increased their activities in online space. Therefore, a great amount of information is shared in online platforms. The users of social networks such as Instagram, Facebook, and Twitter have published noticeable feedbacks about the COVID-19 phenomenon.

As a prominent social network, Twitter allows to analyze social opinions and behaviors. In the past, information retrieval and natural language processing have been used to acquire user opinions on Twitter about things such as natural disasters (Dereli, Eligüzel, & Çetinkaya, 2021; Sreenivasulu & Sridevi, 2020), psychology (Watanabe, Bouazizi, & Ohtsuki, 2018), criminology (Chen, Cho, & Jang, 2015), economics (Bollen, Mao, & Zeng, 2011; Valle-Cruz, Fernandez-Cortez, López-Chau, & Sandoval-Almazán, 2021), politics (Belcastro, Cantini, Marozzo, Talia, & Trunfio, 2020), religion (Nejad, Hosseinzadeh, & Mohammadi, 2018), and medicine (Mahata, Friedrichs, Shah, & Jiang, 2018). Specifically, some studies have been conducted on epidemics such as influenza (Alkouz, Al Aghbari, & Abawajy, 2019), Ebola (Priest & Groves, 2019), and Zika (Lwin et al., 2020), acquiring significant results such as disease prediction models (Alkouz et al., 2019).

During the COVID-19 pandemic, there have been analytical works on the tweets of this subject. When the quarantine was imposed for the first time in India (20–28 of March, 2020) the tweets regarding COVID-19 have been analyzed and the ratio of positive versus negative tweets has been compared (Barkur & Vibha, 2020). Besides, the tweets from US within January to March 2020 have been analyzed concerning social distancing from six facets (e.g. adaptation) and the shares of each facet in each state have been compared (Kwon, Grady, Feliciano, & Fodeh, 2020).

As the COVID-19 pandemic is now the primary global issue, this paper investigates and analyzes the tweets on the subject to extract sentiment trends for major countries associated to this phenomenon. For this purpose, more than two million related English tweets (from March 23 to June 23 of 2020) have been randomly selected and analyzed. At first, by using the GeoNames[1] database, a complete lexicon is collected including the names of locations per nation (name of the country, province's/states, cities) to better retrieve the location that is discussed in the tweets. Note that the conventional methods used for location labelling (such as the geographical tags) are not sufficiently accurate and only consider the locational information of the user. For instance, a user from US may tweet about China, hence, their tweet should fall into China's statistics. The proposed lexicon-based approach, tags the tweets with their content's location. Then, we propose a method based on RoBERTa (Liu et al., 2019) model to analyze the sentimental content of the tweets. RoBERTa has been recently introduced and widely utilized for language model pre-training and has reported promising results (Liu et al., 2019). Lastly, the graphs containing the frequency as well as positive/negative tweets are made for the whole worlds as well as each nation. Analyzing the temporal trend of tweets as well as positive and negative sentiments reveal various implicit information. Finally, a significant correlation between the frequency of COVID-19 related tweets and the official statistics of the daily afflicted cases is detected for most of the countries.

[1] https://www.geonames.org/

The rest of the paper is organized as follows: Section 2 discusses related literature. Section 3 presents the proposed method, and Section 4 discusses the experiments and findings. Finally, Section 5 concludes the results of the study.

## 2. Related work

Today, social networks, especially Twitter, contain a great amount of information shared by users about significant events. The analysis of this information can yield applicable knowledge. First, we need to extract the comments related to a subject and then, based on the application, conduct different processing practices such as location retrieval (Hoang & Mothe, 2018), sentiment analysis, and knowledge discovery. Extracting the positive/negative user opinions about food (Asani, Vahdatnejad, Hosseinabadi, & Sadri, 2020), the travel preferences of tourists (Abbasi-Moud, Vahdat-Nejad, & Mansoor, 2019; Abbasi-Moud, Vahdat-Nejad, & Sadri, 2021), the sentiments of Syrian refugees (Öztürk & Ayvaz, 2018), and food preferences of users (Widener & Li, 2014) are some examples of these processing practices. There have also been various studies analyzing the comments on diseases and epidemics in social networks, as follows.

Psychologists and physicians have acquired useful information by analyzing user opinions about different physical or mental ailments in Twitter. For instance, scholars conducted an analysis of user opinions about different kinds of cancer and chemotherapy (Zhang, Hall, & Bastola, 2018). Besides, the location distribution of users whose tweets indicated depression as well as the effect of different parameters such as education and income on depression have been investigated (Yang & Mu, 2015). Moreover, 184,214 tweets about HPV vaccines have been analyzed, and the frequency graphs of tweets regarding the vaccines as well as the sentiment trends of the tweets are presented (Du, Xu, Song, & Tao, 2017).

User tweets can serve as a tool to predict different epidemics (Velardi, Stilo, Tozzi, & Gesualdo, 2014). In 2015 scholars analyzed 91,495 tweets about H1N1 influenza in India, to extract the hot topics of the tweets (Jain & Kumar, 2015). Moreover 975,752 tweets concerning user opinions about the immediate and long-term symptoms of the Zika pandemic from have been investigated (Khatua & Khatua, 2016). Besides, the frequency of posts in two major social networks (Twitter and Facebook) about Zika Pandemic in Singapore have been compared. The posts were from accounts owned by government, public, corporate bodies, and news agencies (Lwin et al., 2020).

With the COVID-19 pandemic, the user reviews on social networks regarding this phenomena have increased. Therefore, investigating these reviews has attracted researchers. In this regard, the public opinion regarding COVID-19 has been investigated on the Chinese social network of Weibo during January 14 to February 25 of 2020 and the temporal trends of sentiment analysis for hot topics are presented (Zhu, Zheng, Liu, Li, & Wang, 2020). Similarly, relevant posts on Weibo, from January 1 to February 18 of 2020, have been analyzed and the ratio of positive/negative/neutral posts to the total number of posts have been acquired (Wang, Lu, Chow, & Zhu, 2020). Another study has analyzed 178,159 tweets at the last week of February 2020, finding that a larger number of tweets contained positive/neural sentiments compared to negative (Bhat, Qadri, Noor-ul-Asrar Beg, Ahanger, & Agarwal, 2020). Similarly, 24,000 tweets from India during the nationwide lockdown in March 25–28 of 2020 have been investigated, finding that there were more tweets with positive words compared to negative (Barkur & Vibha, 2020).

Finally, 259,529 tweets in the period of January 23 to march 23 of 2020 (with USA geotag) have been investigated regarding different facets of social distancing (negative sentiments, implementation, social disruption, positive sentiments, adaptation, and purpose) (Kwon et al., 2020). The authors found that tweets with negative sentiments were more frequent than those with positive ones.

A review of previous studies reveals that none of them performed sentiment analysis specifically for nations engaged with COVID-19, and besides, mostly used the geotag of a tweet to determine user location which is not accurate; because a user may tweet about another nation.

Because of the importance of the COVID-19 pandemic, this research performs sentiment analysis for the tweets related to COVID-19 in 32 countries with high patient statistics. In contrary to related research that mainly use geotag, we use a complete lexicon of the names of different locations (nation, province/state, and city) for each country to retrieve the location of tweets; because a user may tweet about another country. Finally, the temporal distribution graphs of the positive/negative sentiments for each nation are obtained and investigated.

## 3. Extracting and analyzing Tweets

Twitter has turned into a rich source of user opinions about important phenomena. One of the chief subjects of the tweets in 2020 is COVID-19. User tweets can help extract their sentiments about this phenomenon during a period of time. This research aims to collect and process a large amount of COVID-19 related tweets to acquire the trends of user sentiments over time. First, a large number of COVID-19 related tweets have been collected and their locations have been determined. After that, sentiment analysis is performed on the tweets to identify both positive and negative ones. The proposed method consists of three major steps: collecting the tweets, geotagging them, and performing sentiment analysis.

### 3-1 Collecting tweets

Although COVID-19 pandemic started at the end of 2019, in the first months it only involved China. From February 2020 it started involving other nations and at last, has spread worldwide by the end of March. As this study aims to investigate the effect of COVID-19 pandemic on the world, the tweets have been collected from the late March 2020. Therefore, a large number of tweets from across the globe within the time period of March 23 to June 23 of 2020 have been collected. To extract the target tweets, the keywords of "corona, coronavirus, pandemic, sarscov2, COVID, and COVID19" have been used. The sampling has been performed on the weekly time frame for Tuesdays and a random subset consisting of 150,000 tweets has been selected for each Tuesday.

### 3-2 Location tagging:

To separately analyze the tweets pertaining to each nation, first we need to tag their location. Although previous research studies mainly use the geotag to label the location of a tweet, it could be misleading because a tweet published from a specific country might say about another country. For example, a person in Washington DC may tweet about Canada, and therefore, labelling the location of the tweet solely based on the geotag leads to mislabeling. Therefore, we propose a dictionary-based method for geotagging the tweets. To this end, the GeoNames[2] database is used to gather a list of location names. GeoNames includes geographic information such as the names of countries, states/provinces, cities, postal codes, etc. After that, we extract the names of states/provinces and cities for 32 elected nations[3] (that have the most infected COVID-19 cases), and named the set as "location". The location list (consists of about 7,000 words) is utilized in a GATE (Cunningham, 2002) pipeline as a Gazetteer list, with the tweets being matched with the list. For each instance of a city/province name from a country, the nation's name is attributed as a feature to the tweet.

### 3-3 Sentiment Analysis:

One of the key processes performed on the tweets is sentiment analysis, which is used to gauge the sentimental content of a text. This method is concerned with the attitude, sentiments, and opinions of people about a subject (Medhat, Hassan, & Korashy, 2014). We use sentiment analysis to determine the trend of sentiments and opinions of the people of different nations about COVID-19 during the time.

The sentiment classification model is trained using the RoBERTa (Liu et al., 2019) language model. The RoBERTa model receives a tweet (string) as input and produces a set of token representations $h_t(t = 1, \dots, T)$ as output. As the model is trained using three different datasets, we use a classifier for each dataset (i.e. three classifiers for three datasets).

The input of each classifier is a vector representation (H) of the tweet, which is given by "attention" as follows:

$$H = \sum_t e_t * h_t$$

Wherein

$$e_t = \frac{e^{\alpha_t}}{\sum_j \alpha_j} \qquad \alpha_t = w_{att} h_t$$

In the test phase, the majority vote is utilized to select the final label of the tweet. We make use of three labelled datasets for training the sentiment classifier model: the Stanford sentiment treebank (67,300 samples) (Socher et al., 2013), SemEval 2015 Task 10 (6,800 samples) (Liu et al., 2019), and SemEval 2015 Task 11 (Ghosh et al., 2015) (3,500 samples). To train the sentiment classifier model, the RoBERTa language model with a learning rate of 0.0005 and batch size of 32, is fine-tuned for 12 epochs.

---

2 http://download.geonames.org/export/dump

3 Australia, Belgium, Brazil, Canada, Chile, China, Ecuador, France, Germany, India, Iran, Ireland, Italy, Japan, Mexico, Netherlands, New Zealand, Pakistan, Peru, Portugal, Qatar, Russia, Saudi Arabia, Singapore, South Korea, Spain, Sweden, Switzerland, Turkey, UAE, UK, USA

After calculating the sentiment scores of the tweets, the positive and negative tweets of each week are detected. This provides us with feedback about the trend of public opinion about COVID-19 during the studied period. After that, the frequency of the tweets and the number of positive/negative tweets are computed separately for each of the 32 elected countries (which had the highest statistics of outbreak). Investigating the trend of the graphs reveals interesting information about the public opinions and their correspondence with the spread graph of COVID-19.

## 4. Experiment

A dataset of more than two million tweets related to COVID-19 on the Tuesdays from March 23 to June 23 of 2020, has been collected. To analyze the tweets relevant to each nation, the tweet's location has to be determined. To address this issue, this paper has created a location ontology consisting of 7,000 words of geographical locations via the GeoNames geographical database. This ontology involves the names of states/provinces and cities for 32 countries with the highest statistics of COVID-19 patients.

To extract the location of each tweet, we have used GATE[4], which is an open-source toolkit used to develop any text processing program, and thus is able to process any kind of language (Cunningham et al., 2014). GATE offers many components, including labeling, for different language processing tasks (Cunningham et al., 2014). It performs preprocessing on the tweets to identify the role of the words and therefore, increases labelling accuracy. The location list has been introduced to a GATE pipeline as a Gazetteer list and the tweets are compared to this list. The GATE pipeline extracts the locations stated in the tweets, i.e., for each occurrence of the name of a city, province/state, or country, the name of the corresponding nation is attributed to the tweet as a feature.

Sentiment analysis is a key process on tweets, serving to uncover the positive or negative emotions of a person based on the text. We have employed the RoBERTa (Liu et al., 2019) language model for sentiment analysis. Figure1 shows the frequency of positive and negative tweets as well as the number of confirmed official Covid-19 infected cases obtained from the "John Hopkins university and medicine coronavirus resource center" (available at https://coronavirus.jhu.edu/map.html). It can be seen that in the whole 3-month period, the number of negative tweets is more than the number of positive tweets, which confirms the bad feeling of people about this disease during this period. Besides, the number of negative tweets is roughly ascending; while the number of positive tweets is approximately descending. This fact emphasizes that in the initial stages of the global outbreak, the feelings of people about this epidemic have become worse over time. The official infected cases curve, which has an upward trend, also approves this global trend.

[4] https://gate.ac.uk/

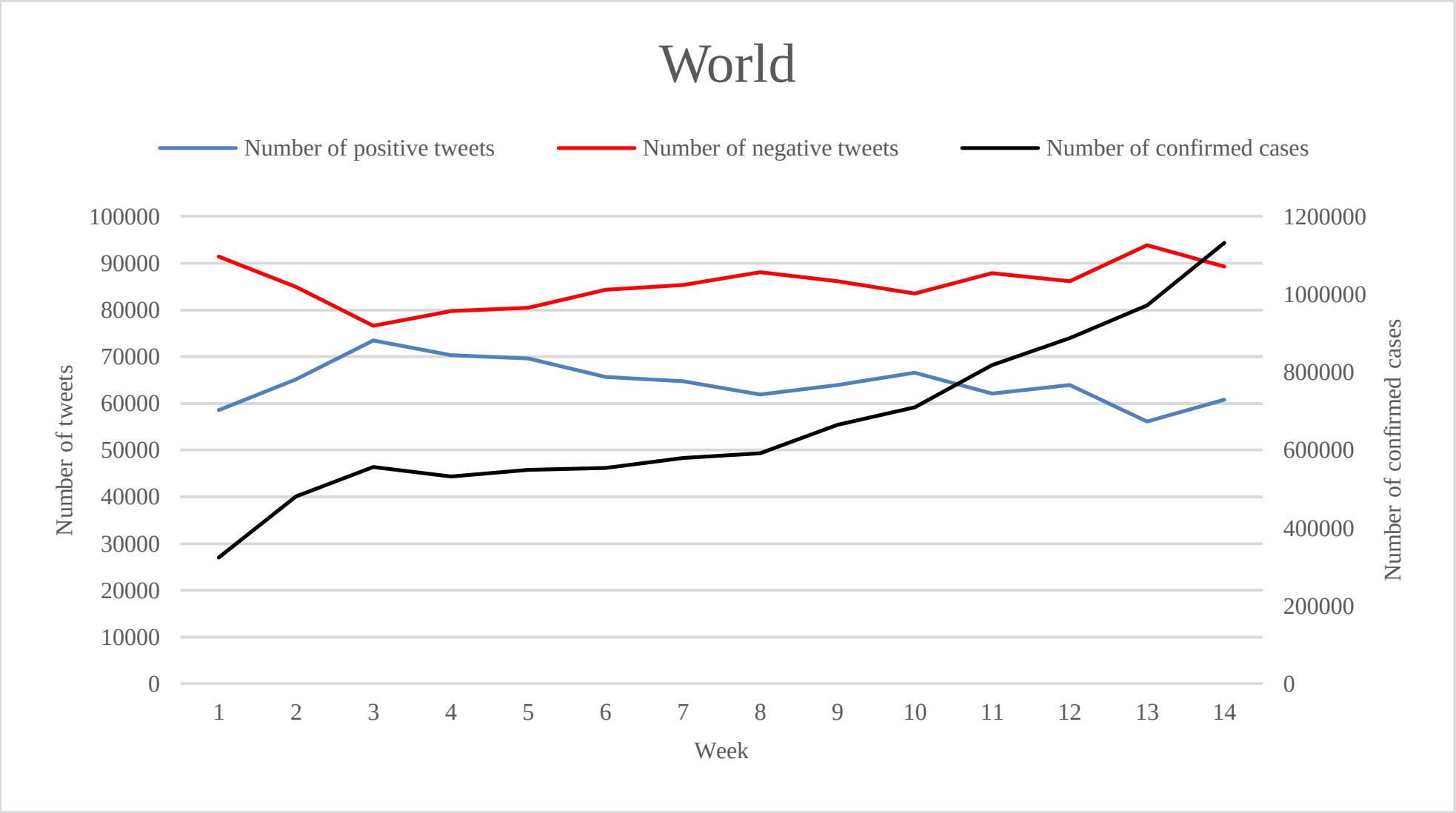

Figure 1. Number of positive and negative tweets and the official statistics of infected cases for the whole world

In the next step, we look through 32 countries with the highest statistics of corona patients. To this end, figure 2 shows the graphs of frequency of tweets and positive/negative sentiments, acquired on a weekly basis (from March 23 to June 23 of 2020). Similar to the previous step, the official statistics of infected cases for each country is also acquired and presented. The left axis of each graph indicates the frequency of the tweets; while the right axis indicates the official infected cases.

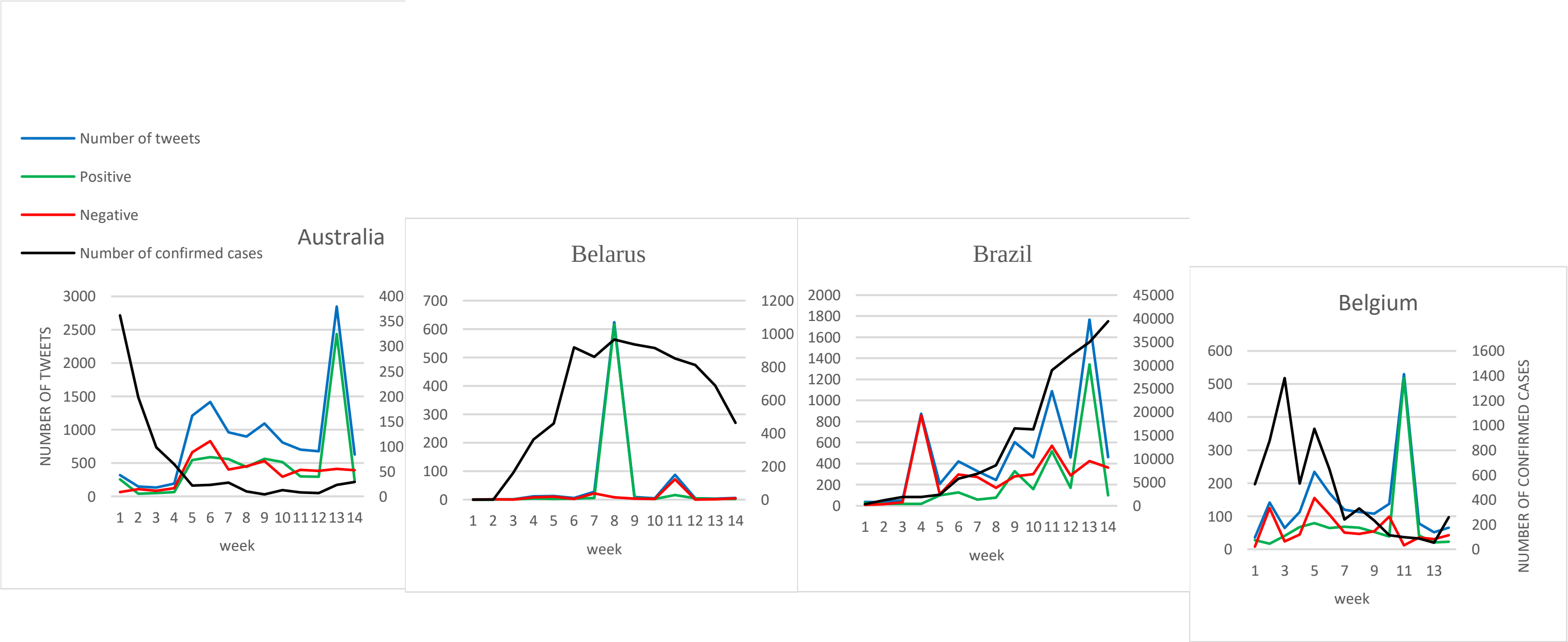

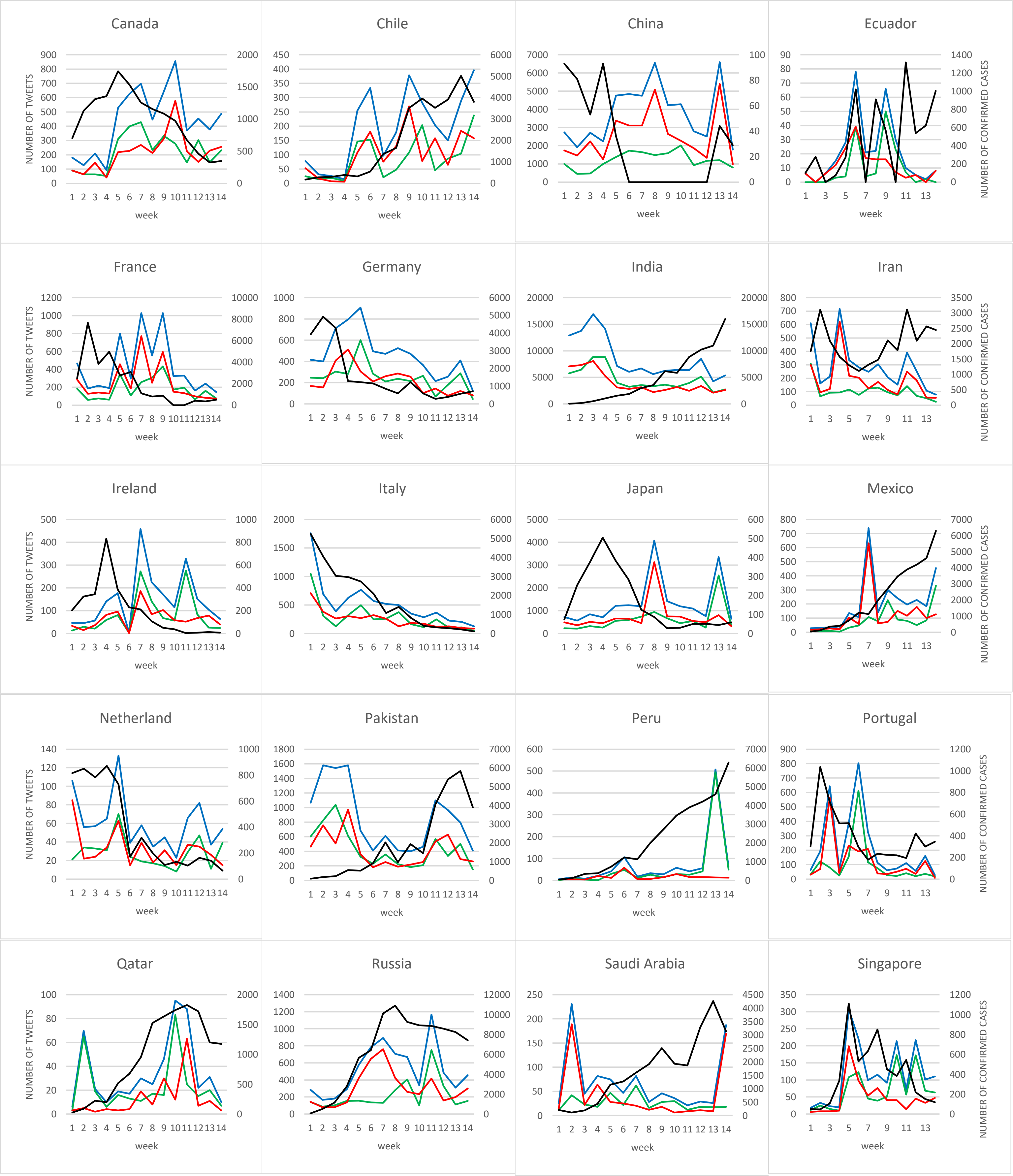

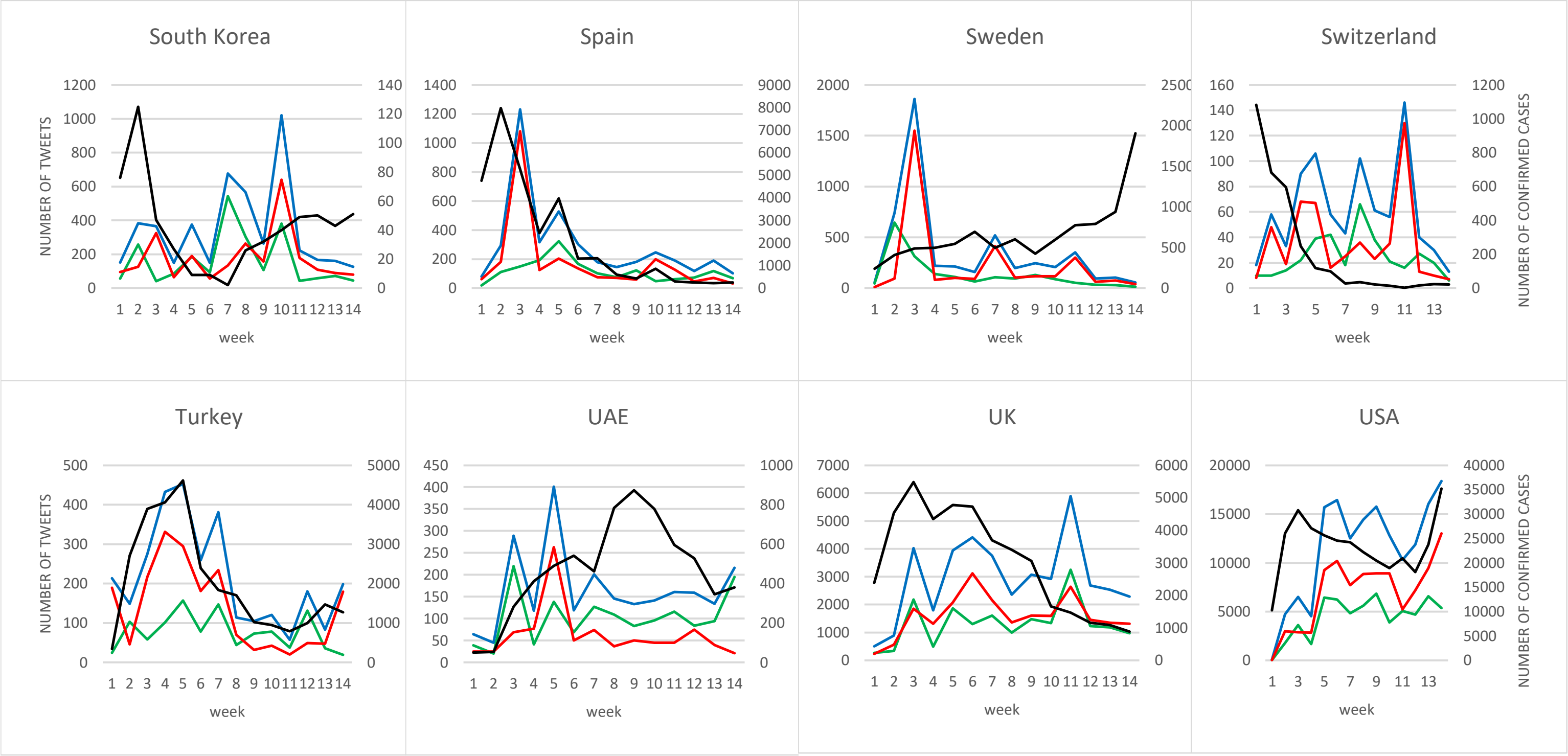

Figure 2. Frequency of tweets, number of positive and negative tweets and the official statistics of infected cases for 32 countries.

Investigating the diagrams reveals that in many countries including Brazil, Belgium, Iran, Germany, Ecuador, Qatar, Italy, Singapore, Turkey, and Spain, there is a correlation between the graphs of the frequency of tweets and the daily registered corona cases. One of the key implications of this correlation is that based on one of the said graphs (tweet frequency or daily registered corona cases), we can predict the future trend of the other graph. For instance, in UAE and Chile, two or three weeks after a maximum in the number of tweets, the statistics of infected cases also peaked.

China is an exception among these countries. Although it has a few official infected cases within the sixth to 12$^{th}$ weeks, this nation is one of the most controversial countries among the users, with many corona-related tweets (mostly with negative sentiment) published about it. Moreover, the pattern of the tweet frequencies is also interesting for India and Pakistan; where users tweeted about the subject several weeks prior to the peak of the pandemic in their countries. The contents of the tweets in these nations during the studied period mostly concern the management of the epidemic, which is understandable considering their large and dense population.

There is a correlation between the graphs of tweets frequency and the total positive/negative tweets in most studied countries. That is, an increase in the total number of tweets also increases the number of positive/negative tweets, while a decrease in this total number also reduces the number of positive/negative tweets. However, there are several exceptions that should be studied more carefully. For example in Portugal, when the number of COVID-19 cases peaked, the negative tweets also peaked; whereas in the following weeks during which the number of confirmed cases reduced, the number of negative tweets also reduced while the number of positive tweets peaked. Finally, by comparing the number of positive tweets versus negative ones, China has the highest ratio of negative tweets. In fact, in all investigated weeks, the number of negative tweets have been more than the number of positive ones. By concentrating on the diagrams of these countries, several other findings are revealed that need much more space to discuss.

## 5. Conclusion:

A method to analyze the COVID-19 related tweets aiming to extract the trend of user sentiments has been presented. The GeoNames database has been utilized to propose a lexicon-based method (developed by the GATE pipeline) for location tagging. Lastly, we have used the RoBERTa model to determine the positive/negative sentiment of each tweet. The frequency graphs for the tweets of different nations has indicated that in most of them, there is a correlation between the peak in the frequency of corona-related tweets and the peak in corona patients. One of the limitations of this study is the sole inclusion of tweets in English language. Moreover, as only tweets on Tuesdays have been sampled, the accuracy of the results may not be high.

Considering that a significant part of the tweets is non-English, a language-independent method can provide a more comprehensive and accurate analysis of the public opinion. Further, considering the natural difference of languages, future works can include non-English ones such as Chinese, Arabic, Farsi, Dutch, Italian, etc. Furthermore, a topic modeling approach can be used to cluster tweets according to their topics and then, analyze each cluster specifically. Similarly, classifying the corona-related tweets according to economic, educational, tourism, psychological, and other contents can help with separately analyzing the tweets in each class based on the special parameters of the subject, thus getting a feedback of COVID-19's effect on said field.